\documentclass[final] {aipproc}

\layoutstyle{8x11double}

\begin{document}

\title{Nonequilibrium Spin-transfer Torque in SFNFS Junctions}



\author{Erhai Zhao and J. A. Sauls}{
  address={Department of Physics and Astronomy, Northwestern University, Evanston, IL 60208, USA}
}

\begin{abstract}
We report theoretical results for the nonequilibrium spin current
and spin-transfer torque in voltage-biased SFNFS Josephson
structures. The subharmonic gap structures and high voltage
asymptotic behaviors of the dc and ac components of the spin current
are analyzed and related to the spin-dependent inelastic scattering
of quasiparticles at both F layers.
\end{abstract}

\maketitle

Spin-polarized current passing through a ferromagnet can transfer
spin angular momentum to the ferromagnet and exert a torque on the
magnetic moment \cite{slon,berger}. This mechanism offers unique
opportunities to manipulate the magnetic state of nanomagnets.
Experimentally spin-transfer torque driven magnetization precession
and magnetization reversal have been observed in magnetic
multilayers \cite{tsoi,nature03}. A well studied multilayer system
is the magnetic nanopillar which consists of a ferromagnet-normal
metal-ferromagnet (FNF) trilayer connected to normal metal
electrodes. The typical thickness of each layer is several nm, and
the diameter of the pillar is of the order of 50 nm \cite{nature03}.
When sandwiched between superconducting electrodes, the FNF trilayer
can mediate finite Josephson coupling to form a SFNFS Josephson
junction \cite{bell}. In such junctions, scattering of quasiparticle
at the magnetic interfaces is sensitive to the phase shift in each F
layer as well as the condensate phase difference $\phi$ across the
junction. This indicates the spin momentum transfer between the
quasiparticles and ferromagnets can be tuned by varying $\phi$.
Waintal and Brouwer calculated the phase-sensitive equilibrium
torque in SFNFS junctions \cite{waintal2}. They also showed the
nonequilibrium torque in NFNFS junctions acquires novel features,
e.g. it can favor perpendicular configuration of the two moments
\cite{waintal1}.

\begin{figure}
  \includegraphics[width=2.2in]{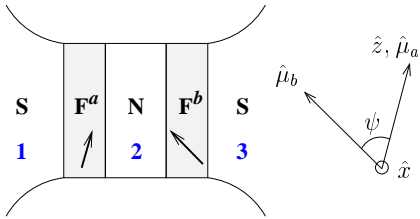}
  \caption{Schematic of SFNFS junction. The magnetization
 $\vec{\mu}_a$ is along the $z$ axis (the quantization axis for spin), while $\vec{\mu}_b$ is at
 polar angle $\psi$ in the $yz$ plane. The $x$ axis is along $\vec{\mu}_a\times
 \vec{\mu}_{b}$. Notice our choice of axis differs from Ref. \cite{waintal2}.}
  \label{fig1}
\end{figure}

Here we investigate the nonequilibrium spin-transfer torque in
SFNFS point contacts under bias voltage $V$. The setup and the
coordinate system are shown in Fig. \ref{fig1}. The two F layers
are labelled by indices $a$ and $b$, respectively. The N spacer is
assumed to be transparent and in the clean limit. For simplicity
we model each F layer as a delta function barrier with
spin-dependent transmission probability $D_{\uparrow}\neq
D_{\downarrow}$, so the spin mixing angle \cite{zls}
$\vartheta=\arcsin\sqrt{D_{\uparrow}}-\arcsin\sqrt{D_{\downarrow}}$.
The scattering matrix of F$^a$ and F$^b$ are related by spin
rotation of angle $\psi$. Since the condensate phase difference
evolves at Josephson frequency $\omega_J =2eV/\hbar$, the spin
current in each region, $\mathbf{I}_{i}$ ($i=1,2,3$), also
oscillates with time. For example,
\begin{equation}
\mathbf{I}_2(t)=\mathbf{I}_0+\sum_{k=1}^{\infty} [\mathbf{I}_{k,c}
\cos (k\omega_J t)+\mathbf{I}_{k,s}\sin (k\omega_J t)].
\label{eq1}
\end{equation}
We use the quasiclassical theory of superconductivity \cite{rs}
and solve numerically the transport equations for the Green's
functions in each region which obey proper boundary conditions at
interface F$^a$ and F$^b$ \cite{zls}. The spin currents are then
determined from the local Keldysh Green's functions. Within this
approach the proximity effect and the multiple Andreev reflection
(MAR) at both interfaces are fully taken into account.

\begin{figure}
  \includegraphics[width=2.8in]{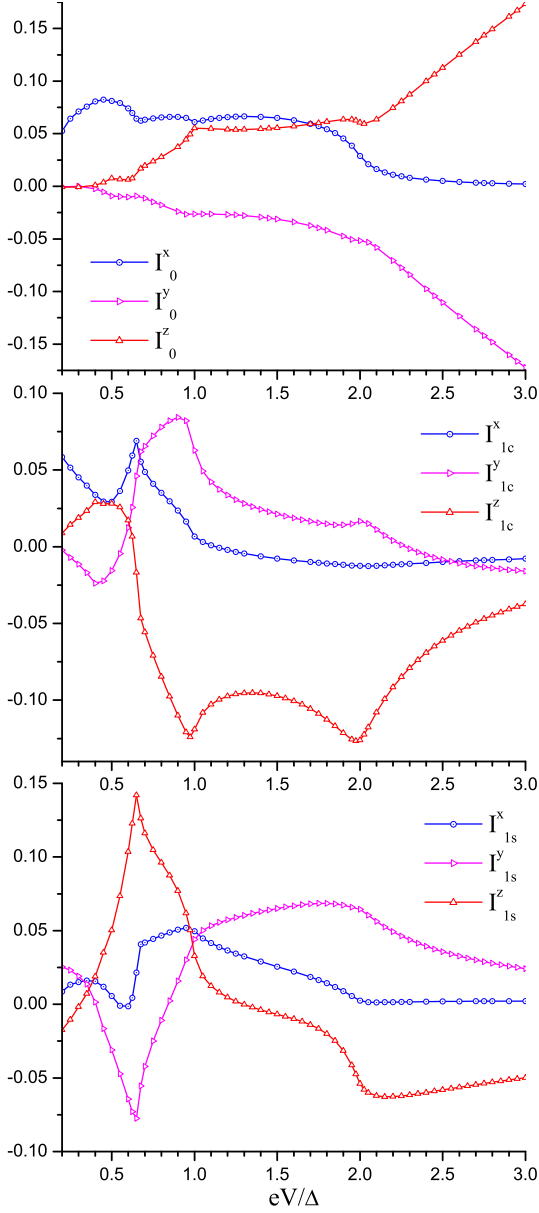}
  \caption{Dc (top panel), cosine (middle panel) and sine (bottom panel)
  part of the first Fourier components of the
  spin current in the N layer. Each panel features the spin current along $x$,
  $y$, and $z$ direction. $D_{\uparrow}=0.81$, $D_{\downarrow}=0.64$, $\psi=\pi/2$.}
  \label{fig2}
\end{figure}

The spin current in the N layer, measured in unit of $N_fv_f\Delta
A\hbar/2$ where $A$ is the contact cross sectional area and $\Delta$
is the superconducting gap, is shown in Fig. \ref{fig2} for
$\psi=\pi/2$ and zero temperature. As in the case of nanopillars
with normal metal electrodes, the dc spin current flow is a
consequence of spin filtering at the F layers. Its voltage
dependence, however, is nonlinear and possesses subharmonic gap
structures due to the onset and resonances of MAR processes. At
$V=0$, the dc spin current vector is along $x$ direction, which
tends to cause the moments to precess around each other
\cite{waintal2}. As $V$ is increased, the magnitude of dc currents
along $y$ and $z$ direction, which tend to bring the moments towards
or away from each other, grow with $V$. The ac spin current
originates from the interference between MAR processes of different
order. Its magnitude changes rapidly at voltages below $2\Delta$ and
decays to zero at high voltages.

\begin{figure}
  \includegraphics[width=2.8in]{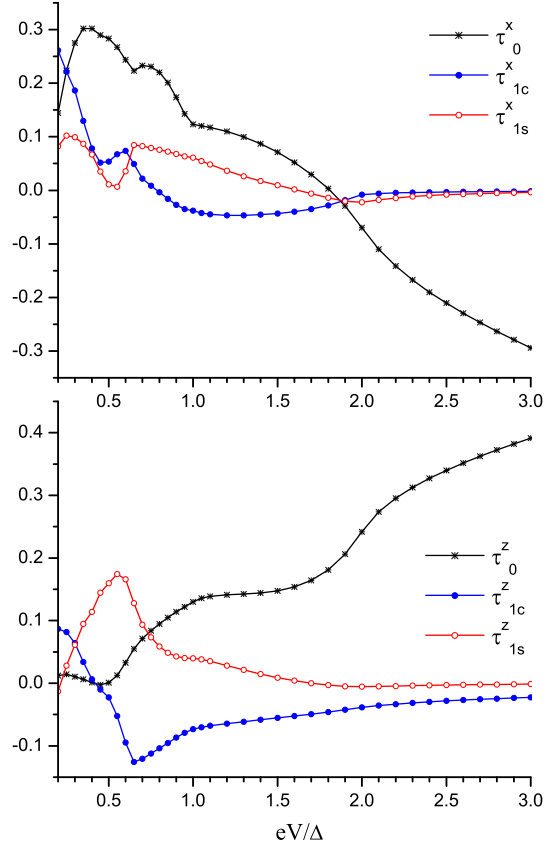}
  \caption{Spin-transfer torque on F$^b$ in $x$ (upper panel) and $z$ (lower panel)
  direction. $D_{\uparrow}=0.95$, $D_{\downarrow}=0.6$, $\psi=\pi/2$.
  } \label{fig3}
\end{figure}

The spin-transfer torque on F$^a$ is given by
$\vec{\tau}_a(t)=\mathbf{I}_1(t)-\mathbf{I}_2(t)$, and the torque on
F$^b$ is given by $\vec{\tau}_b(t)=\mathbf{I}_2(t)-\mathbf{I}_3(t)$.
It is convenient to expand $\vec{\tau}_{a/b}(t)$ in Fourier series
similar to Eq. (\ref{eq1}). Fig. \ref{fig3} shows the $k=0$ and
$k=1$ components of $\vec{\tau}_b$, in unit of $N_fv_f\Delta
A\hbar/2$, as functions of the bias voltage at $T=0$. Notice
$\vec{\tau}_b$ is perpendicular to $\vec{\mu}_b$ and lies in the
$xz$ plane ($\psi=\pi/2$). An interesting feature of the dc torque
along $x$ direction is that it changes sign around voltage
$1.8\Delta/e$. The magnitude of the dc torque becomes linear to $V$
at high voltages. The ac torque becomes vanishingly small compared
to the dc torque for $V>2\Delta$ because the number of subgap
Andreev reflections, which constitute the dominant contributions to
the ac spin current, is inversely proportional to $V$.

\end{document}